\documentclass[aps,prl,reprint,twocolumn,superscriptaddress]{revtex4-2}

\usepackage[T1]{fontenc}
\usepackage{lmodern}
\usepackage{microtype}
\usepackage{amsmath,amssymb,mathtools,bm}
\usepackage{siunitx}
\usepackage{hyperref}
\usepackage{cleveref}

\hypersetup{colorlinks=true,linkcolor=blue,citecolor=blue,urlcolor=blue}
\begin{document}

\title{Planar Interfaces for Transmission of Chiral Spin Textures}
	
\author{Robin Msiska}
\affiliation{Theoretical Division, Los Alamos National Laboratory, Los Alamos, NM 87545, USA}
\affiliation{Center for Memory and Recording Research, University of California, San Diego, La Jolla, CA 92093, USA}
\author{Cynthia J.O. Reichhardt}
\affiliation{Theoretical Division, Los Alamos National Laboratory, Los Alamos, NM 87545, USA}
\author{Charles Reichhardt}
\affiliation{Theoretical Division, Los Alamos National Laboratory, Los Alamos, NM 87545, USA}
\author{Eric Fullerton}
\affiliation{Center for Memory and Recording Research, University of California, San Diego, La Jolla, CA 92093, USA}
\author{Avadh Saxena}
\affiliation{Theoretical Division, Los Alamos National Laboratory, Los Alamos, NM 87545, USA}

\begin{abstract}
Lateral magnetic interfaces provide a direct way to test whether skyrmions remain robust when driven across abrupt changes in material parameters and magnetic order. Here we study skyrmion transmission across planar ferromagnet--ferromagnet (FM--FM), antiferromagnet--antiferromagnet (AFM--AFM), ferromagnet--antiferromagnet (FM--AFM), and antiferromagnet--ferromagnet (AFM--FM) interfaces using micromagnetic simulations and analytic reduced-coordinate criteria. The outcomes are organized into phase diagrams according to the morphology formed in the receiving region, distinguishing compact transmission from deformed skyrmions, stripe-domain states, amorphous textures, and relaxation into the background. Same-order FM--FM and AFM--AFM skyrmion transmission is captured by an analytically defined range of the reduced Dzyaloshinskii--Moriya interaction, identifying the wall-softening regime that supports compact transmission without stripe formation. Mixed-order FM--AFM and AFM--FM interfaces are directionally distinct, requiring conversion between ferromagnetic magnetization and antiferromagnetic N\'eel textures. These results show that planar interfaces act as active transport elements and provide reduced design criteria for heterogeneous skyrmion tracks.
\end{abstract}

\date{\today}

\maketitle

\section*{Introduction}

\begin{figure}[t]
    \includegraphics[scale=0.38]{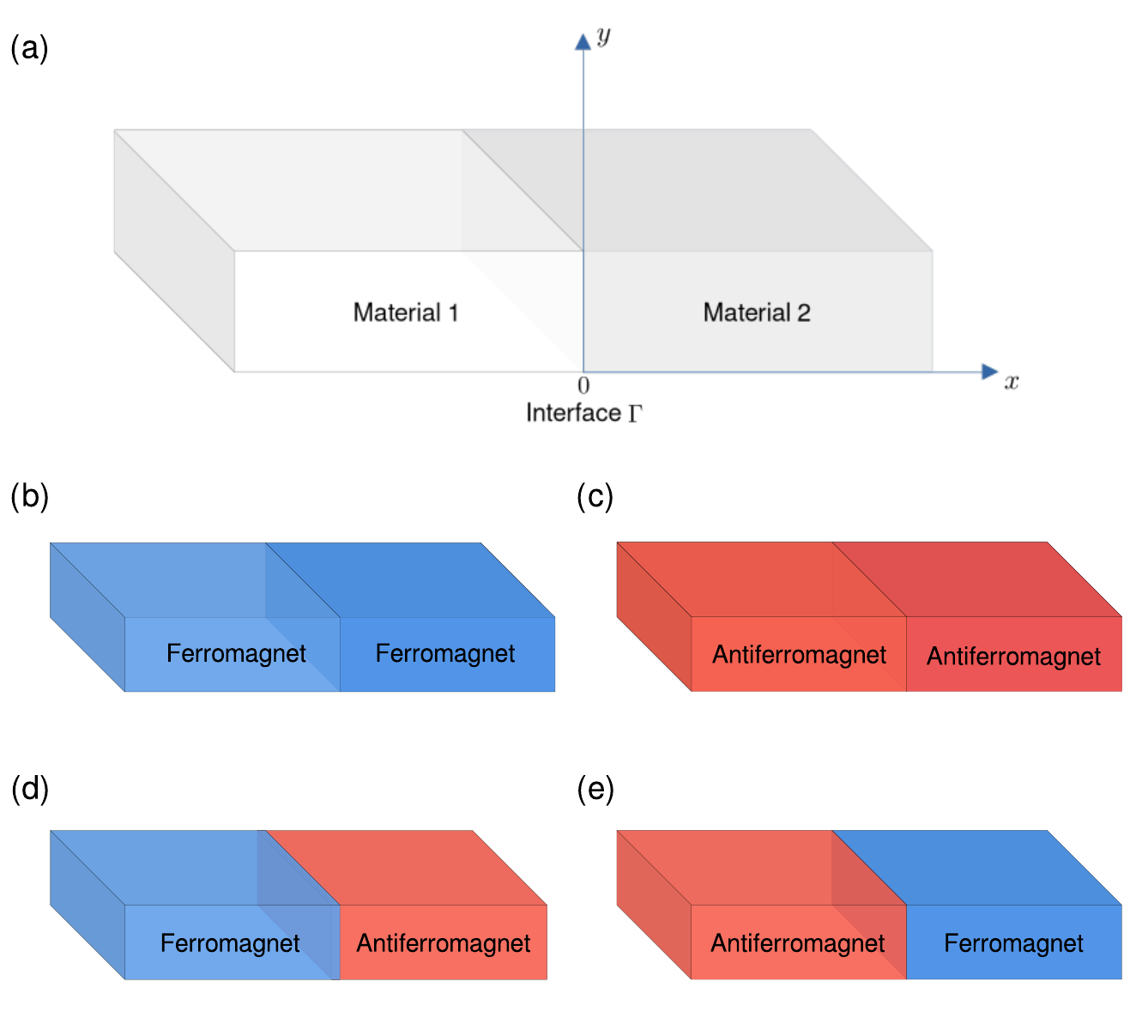}
    \caption{\label{fig:interface_schematic}
Planar interface geometries considered in this work.
(a) General source--receiving geometry, with a skyrmion driven from material 1 on the left toward an interface \(\Gamma\) separating two magnetic media, and material 2 on the right taken as the receiving region.
(b)--(e) Block representations of the four interface classes, with attached colored regions indicating the two media on either side of the interface. Blue denotes a ferromagnetic region and red denotes an antiferromagnetic region.
(b) FM--FM interface, where the magnetic order type is ferromagnetic on both sides.
(c) AFM--AFM interface, where the magnetic order type is antiferromagnetic on both sides.
(d) FM--AFM interface, where a ferromagnetic source region is joined to an antiferromagnetic receiving region.
(e) AFM--FM interface, where an antiferromagnetic source region is joined to a ferromagnetic receiving region.
The ordering of each label denotes the direction of traversal from source to receiving region.
}
\end{figure}

Magnetic skyrmions have emerged as one of the most compelling quasiparticles in spintronics because they combine nanoscale dimensions, topological structure and current-driven mobility in a form that is directly relevant to memory and logic concepts~\cite{Bogdanov1989, Fert2013,Nagaosa2013,Sampaio2013,Iwasaki2013,Back2020}. Over the past decade, theoretical and experimental advances have shown that skyrmions can be stabilized in chiral magnetic films and multilayers, displaced with electrical currents, and manipulated in confined geometries, making them credible candidates for device-level information transport~\cite{Rosch2013,Tomasello2014,Zhang2015,EverschorSitte2018,Legrand2020,Zhang2020}. As the field moves from proof-of-principle demonstrations toward functional circuitry, however, the central question is no longer simply whether a skyrmion can exist in a homogeneous medium. The more demanding question is whether it can remain a compact, functionally identifiable object while traversing the heterogeneous landscapes that realistic architectures inevitably contain.

That shift in emphasis makes interfaces fundamental. In any practical skyrmion device, the regions optimized for nucleation, propagation, storage, sensing, or deletion need not coincide, and may in fact favor very different micromagnetic parameter sets. Internal boundaries may therefore not be incidental defects in an otherwise uniform medium, but natural consequences of device design. A skyrmion crossing such a boundary is forced to adapt, while already in motion, to a new balance of exchange stiffness, anisotropy, Dzyaloshinskii--Moriya interaction (DMI), and interfacial torque. Boundary crossing thus probes a stronger and more operational notion of robustness than bulk metastability alone. That is, a skyrmion that is perfectly viable in a uniform film may still fail as an information carrier if the deformation required for transmission is too severe~\cite{Rohart2013,Raeliarijaona2018,Menezes2019}. This consideration motivates the minimal geometry studied here, in which a skyrmion is driven from one magnetic region toward a planar interface and into a second region, as summarized in Figure~\ref{fig:interface_schematic}. The incident side is treated as the source region and the far side as the receiving region.

Related studies have already shown that skyrmions can interact strongly with material steps and interfaces, either by moving through them~\cite{Zhou2019,Littlehales2024} or by propagating along them as guided interfacial objects~\cite{Iwasaki2014,Martinez2018,Raab2024}. These examples demonstrate that boundaries are not merely passive obstacles, but can redirect, confine, or reshape skyrmion motion. What remains less settled is how the outcome changes when the interface also separates distinct magnetic orders, so that transmission requires not only deformation of the texture but, in mixed-order cases, adaptation to different order-parameters.

Experimentally, such planar magnetic-order interfaces could be approached in several ways, including gated van der Waals magnets~\cite{Huang2018,Jiang2018}, chemically ordered films locally modified by ion irradiation~\cite{Maat2003,Heidarian2015,Bennett2018,Griggs2020}, irradiation-induced conversion between ferromagnetic and antiferromagnetic order in layered magnets such as CrSBr~\cite{Long2024}, moir\'e-patterned magnetic heterostructures~\cite{Song2021,Xu2022,Huang2023}, and phase-change materials such as FeRh~\cite{Maat2005,Uhlir2016}. Direct-write laser processing provides an additional route by locally modifying magnetic anisotropy, compensation temperature, or exchange coupling with spatially varying annealing doses~\cite{Riddiford2025,Brock2025}. This approach is particularly relevant to rare-earth--transition-metal ferrimagnets, in which skyrmions have already been experimentally stabilized and imaged~\cite{Woo2018,Luo2023}. In these platforms, the interface need not correspond to a physical edge between two deposited films; it may instead be written by electrostatic control, structural disorder, twist-controlled exchange, or local modification of the magnetic phase. These possibilities motivate treating the planar boundary itself as a controllable element rather than as a passive junction.

Within this geometry, we consider interfaces between ferromagnetic (FM) and antiferromagnetic (AFM) media, giving four cases in total: FM--FM, AFM--AFM, FM--AFM and AFM--FM. The first two preserve the magnetic order across the boundary, whereas in the latter two the order changes across the interface, with the ordering indicating the direction of traversal. This distinction is substantive rather than merely notational. In a FM, the skyrmion is a texture of a single magnetization field and generally exhibits a nonzero transverse response under current drive, namely the skyrmion Hall effect~\cite{Everschor2011,Schulz2012,Litzius2017,Jiang2017}. In an AFM, by contrast, the relevant topological object is most naturally described as a texture of the N\'eel field built from two coupled sublattice textures, and compensation of the opposite sublattice Magnus forces strongly alters the dynamics and suppresses the conventional Hall response in the ideal limit~\cite{BarkerTretiakov2016,Zhang2016,Msiska2022}. Reversing a mixed-order interface therefore does more than relabel the boundary. It exchanges source and receiving media across an interface at which not only the material parameters but also the character of the underlying order parameter may change, so that the problem is one of interfacial transmission in the strongest sense.

Prior literature on planar interfaces has focused mainly on the propagation of spin waves and magnons. In FM--AFM interfece systems, continuum models have been used to derive interfacial matching conditions for the FM magnetization and the AFM sublattice magnetizations, including the effects of finite-thickness transition regions. These conditions have been applied to describe spin-wave reflection, transmission, phase shifts, and evanescent modes~\cite{Busel2018,Busel2019,Busel2021}. Extensions to AFM--AFM interfaces and magnonic crystals have further addressed nonlinear and supercritical propagation across compensated and uncompensated multisublattice boundaries~\cite{Gorobets2023,Gorobets2024}. Unlike these extended or resonant excitations, we consider magnetic textures with spatially confined internal structures that can deform, become pinned, or transform as they traverse a planar interface.

Against this background, the purpose of the present work is to establish a unified map of skyrmion transmission outcomes across these four interface classes. Rather than restricting the analysis to a binary success--failure measure, the study resolves the quality of the transmitted state, distinguishing compact skyrmion reception from deformed, stripe-like, amorphous or nontransmitted outcomes. The goal is to identify how those outcomes are organized by the micromagnetic control parameters and by the drive, and to condense the resulting phase maps into reduced variables that have direct physical meaning on the receiving side, including the effective anisotropy, the critical DMI scale and the corresponding dimensionless chirality ratio. In that sense, the paper is intended not only as a survey of interface-dependent transmission, but as a step toward design rules for heterogeneous skyrmion circuitry, where transmission, conversion and failure can each be used deliberately rather than encountered accidentally. The manuscript is organized as follows: We first develop the analytic model and introduce the reduced criteria used to interpret skyrmion stability and transmission. We then present the simulation results and transmission maps for the four interface classes, followed by a discussion of their physical implications and a concluding summary of the principal findings.

\section{Model}

We describe the magnetic system within a continuum micromagnetic model. We consider an ultrathin magnetic film in the \(xy\) plane containing a sharp planar interface
\begin{equation}
\Gamma=\{x=0\},
\end{equation}
which separates two regions,
\begin{equation}
\Omega_L=\{x<0\},\qquad \Omega_R=\{x>0\},
\end{equation}
as shown in Figure~\ref{fig:interface_schematic}. The left region \(\Omega_L\) is denoted material 1, and the right region \(\Omega_R\) is denoted material 2. Material parameters are taken to be piecewise constant across the interface. This is the natural continuum description of an abrupt compositional or lithographically defined boundary, and it provides the minimal setting in which to ask whether a skyrmion can survive, deform, or convert as it crosses between distinct magnetic environments.

The time evolution of each unit magnetization field $\bm m$ is governed by the Landau--Lifshitz--Gilbert equation~\cite{Landau1935} supplemented by an external driving torque $\bm\tau$~\cite{Slonczewski1996},
\begin{equation}
\frac{\partial \bm m}{\partial t}=-\gamma
\bm m\times\bm H_{\mathrm{eff}}
+
\alpha
\bm m\times
\frac{\partial\bm m}{\partial t}
+
\bm\tau,
\label{eq:LLG_general}
\end{equation}
where $\gamma$ is the gyromagnetic ratio and $\alpha$ is the Gilbert damping parameter. The effective field is obtained variationally from the total micromagnetic energy $E$ and expressed as
\begin{equation}
\bm H_\mathrm{eff}
=-\frac{1}{M_s}
\frac{\delta E}{\delta\bm m},
\label{eq:effective_field_general}
\end{equation}
with the film thickness understood to be absorbed into the two-dimensional energy functional.

In this study, we model the torque term $\bm \tau$ using the Zhang--Li formulation of spin-transfer torque (STT)~\cite{ZhangLi2004}, retaining both its adiabatic and nonadiabatic components
\begin{equation}
\bm\tau
=-\left(\bm u\cdot\nabla\right)\bm m
+
\beta
\bm m\times
\left[
\left(\bm u\cdot\nabla\right)\bm m
\right].
\label{eq:STT_general}
\end{equation}
Here $\bm u$ is the spin-drift velocity associated with the applied electrical current density $\bm j$, described by
\begin{equation}
\bm u
= \frac{\mu_B P}{eM_s(1+\beta^2)}\bm j,
\label{eq:spin_drift_velocity}
\end{equation}
where the parameter $\beta$ characterizes the nonadiabatic contribution to the STT, $P$ is the dimensionless current spin polarization, $\mu_B$ is the Bohr magneton, $e>0$ is the elementary-charge magnitude, $g$ is the electronic Land\'e factor, and $M_s$ is the saturation magnetization.

To encompass both same-order and mixed-order interfaces, we express the total system energy in the generic form
\begin{align}
E[\bm q_L,\bm q_R]
= 
\int_{\Omega_L}\!d^2r\,\mathcal E_L[\bm q_L]
+
\int_{\Omega_R}\!d^2r\,\mathcal E_R[\bm q_R]
\\ +
\int_{\Gamma}\!ds\,w_\Gamma(\bm q_L,\bm q_R), \nonumber
\label{eq:Etot_general}
\end{align}
where $\mathcal E_L$ and $\mathcal E_R$ represent the energy densities of the two adjoining left ($L$) and right ($R$) media, respectively, while $\bm q_L$ and $\bm q_R$ are the associated order parameters, and $w_\Gamma$ is an explicit interface energy density. The latter is optional for a plain material step within a single magnetic order, but essential once the two sides are coupled through distinct order parameters, as in FM--AFM or AFM--FM interfaces~\cite{Kiwi2001}.

Stationary configurations satisfy $\delta E=0$ under admissible variations tangent to the relevant constraint manifold. In all cases considered below, the corresponding Euler--Lagrange equations are accompanied by boundary terms localized on $\Gamma$, so that the interface enters as a local torque-balance condition rather than as a purely geometric divider. Choosing a unit normal $\hat{\bm n}$ directed from left to right, the interface conditions may be written schematically as
\begin{equation}
\bm J_L
+
\Big(\frac{\delta w_\Gamma}{\delta \bm q_L}\Big)_{\!\perp}
=0,
\qquad
\bm J_R
-
\Big(\frac{\delta w_\Gamma}{\delta \bm q_R}\Big)_{\!\perp}
=0,
\label{eq:general_interface_balance}
\end{equation}
where $\bm J_L$ and $\bm J_R$ are the boundary fluxes generated by the corresponding bulk energies and the subscript $\perp$ denotes projection onto the tangent space of the constrained order parameter. Equation~\eqref{eq:general_interface_balance} is the common variational backbone of all four interface classes and its detailed derivation is presented in Appendix~\ref{app:interface_variation}.

\subsection{FM--FM interfaces}

For a same-order FM boundary, both sides are described by a unit magnetization field $\bm m(\bm r)$ with $|\bm m|=1$. In an ultrathin film with perpendicular anisotropy and interfacial DMI, the standard micromagnetic energy density is
\begin{equation}
\mathcal E_{\rm FM} =
A|\nabla\bm m|^2
+K_{\mathrm{eff}}(1-m_z^2)
+\mathcal E_{\mathrm{iDMI}},
\label{eq:bulk_density_FM}
\end{equation}
with
\begin{equation}
K_{\mathrm{eff}}
=
K_u-\frac{\mu_0 M_s^2}{2},
\label{eq:Keff_FM}
\end{equation}
and
\begin{equation}
\mathcal E_{\mathrm{iDMI}}
=
D\Big[m_z(\nabla\!\cdot\!\bm m)-(\bm m\!\cdot\!\nabla)m_z\Big].
\label{eq:iDMI_FM}
\end{equation}
Here $A$ denotes the exchange stiffness, $D$ the interfacial DMI strength constant, $M_s$ the saturation magnetization, $K_u$ the uniaxial perpendicular anisotropy, and $K_{\mathrm{eff}}$ the corresponding effective anisotropy including the demagnetizing contribution~\cite{Thiaville2012,Rohart2013,Abert2019}. The interfacial DMI favors homochiral domain walls with radial in-plane magnetization and consequently stabilizes N\'eel skyrmions. This is the standard continuum model for N\'eel skyrmions in ultrathin FMs~\cite{Bogdanov2001,Kiselev2011,Rohart2013}.

Variation of the exchange interaction term gives the familiar boundary flux $2A\,\partial_n\bm m$. The DMI contribution supplies an additional chiral boundary torque, so that the natural free-edge condition takes the form
\begin{equation}
2A\,\partial_n\bm m
+
D\,(\hat{\bm z}\times\hat{\bm n})\times\bm m
=0.
\label{eq:free_edge_bc_FM}
\end{equation}
For an abrupt FM--FM interface with no explicit interfacial term, $w_\Gamma=0$, the two boundary terms combine to give the projected jump condition
\begin{equation}
\Big[
2A\,\partial_n\bm m
+
D\,(\hat{\bm z}\times\hat{\bm n})\times\bm m
\Big]_L^R
=
0,
\label{eq:FMFM_jump}
\end{equation}
where $[X]_L^R:= X_R-X_L$. The exchange interaction therefore tries to smooth the texture across the boundary, while the DMI discontinuity changes the preferred chiral twist. The normal derivative adjusts so that the total flux remains continuous. This is the basic analytic kernel behind interface-induced twists, skyrmion deflection, and pinning at FM material steps~\cite{Mulkers2017,Mulkers2018,Raeliarijaona2018,Menezes2019}.

\subsection{AFM--AFM interfaces}

For an AFM region we work in a two-sublattice description, introducing unit vectors $\bm m_1$ and $\bm m_2$ on the two sublattices. A minimal continuum energy density is
\begin{align}
\mathcal E_{\rm AFM}
=&
A\sum_{a=1,2}|\nabla\bm m_a|^2
+
\mathcal J_{\rm AFM}\,\bm m_1\!\cdot\!\bm m_2
\\[10pt] &+
K\sum_{a=1,2}(1-m_{a,z}^2)
+
\sum_{a=1,2}\mathcal E_{{\rm iDMI},a},
\label{eq:bulk_density_AFM}
\end{align}
with $\mathcal J_{\rm AFM}>0$ enforcing antiparallel alignment and with $\mathcal E_{{\rm iDMI},a}$ of the same interfacial form as in Equation~\eqref{eq:iDMI_FM}, applied to each sublattice field indexed by $a$~\cite{BarkerTretiakov2016,Zhang2016}. In the strong-exchange limit it is often convenient to introduce the N\'eel vector and canting field, respectively as
\begin{equation}
\bm n=\frac{\bm m_1-\bm m_2}{2},
\qquad
\bm l=\frac{\bm m_1+\bm m_2}{2},
\label{eq:n_l_def}
\end{equation}
with $|\bm l|\ll1$ and $\bm n^2+\bm l^2=1$. The present subsection, however, retains the sublattice-resolved form because it generalizes most directly to mixed-order interfaces.

Variation now yields one bulk equation and one boundary flux for each sublattice. The corresponding interface fluxes are
\begin{equation}
\bm J_{\rm AFM}
=
2A\,\partial_n\bm m_a
+
D_a\,(\hat{\bm z}\times\hat{\bm n})\times\bm m_a,
\qquad {\rm for}~a=1,2
\label{eq:AFM_flux}
\end{equation}
where $D_a$ allows, in principle, for sublattice-dependent chiral couplings. For a plain AFM--AFM material step with no additional interface term,
\begin{equation}
[\bm J_1]_L^R=0,
\qquad
[\bm J_2]_L^R=0.
\label{eq:AFMAFM_jump}
\end{equation}
Thus, as in the FM case, the interface acts as a local torque-balance condition, but now it must be satisfied simultaneously by both sublattices. In the strong-exchange limit these two conditions may be re-expressed in terms of the N\'eel field $\bm n$, yielding the same structural conclusion: an AFM skyrmion crosses the boundary only if the staggered texture can accommodate the interfacial twist while preserving sublattice locking.

\subsection{FM--AFM and AFM--FM interfaces}

For mixed-order interfaces, the two sides are not described by the same order parameter. A FM region carries a single magnetization field $\bm m$, whereas an AFM region carries two coupled sublattice fields $\bm m_1$ and $\bm m_2$, or the N\'eel field $\bm n$. As a result, the interface cannot be treated as a simple material step with $w_\Gamma=0$. An explicit interfacial coupling is required.

For definiteness, consider first an FM--AFM interface, with the FM on the left and the AFM on the right. A minimal sublattice-resolved interface energy is
\begin{equation}
w_\Gamma
=
-
J_{\Gamma 1}\,\bm m\!\cdot\!\bm m_1
-
J_{\Gamma 2}\,\bm m\!\cdot\!\bm m_2
+
w_\Gamma^{\rm ani},
\label{eq:wGamma_mixed}
\end{equation}
where $J_{\Gamma 1}$ and $J_{\Gamma 2}$ are interface exchange couplings to the two AFM sublattices and $w_\Gamma^{\rm ani}$ denotes any additional interface anisotropy or exchange-bias-like term~\cite{Kiwi2001}. Introducing the combinations
\begin{equation}
J_\Gamma^{(\pm)}=\frac{J_{\Gamma 1}\pm J_{\Gamma 2}}{2},
\label{eq:Jpm}
\end{equation}
and using $\bm m_{1,2}=\bm l\pm \bm n$, Equation~\eqref{eq:wGamma_mixed} may be rewritten as
\begin{equation}
w_\Gamma
=
-2J_\Gamma^{(+)}\,\bm m\!\cdot\!\bm l
-2J_\Gamma^{(-)}\,\bm m\!\cdot\!\bm n
+
w_\Gamma^{\rm ani}.
\label{eq:wGamma_nl}
\end{equation}
Equation~\eqref{eq:wGamma_nl} makes the mixed-order physics transparent. The symmetric combination $J_\Gamma^{(+)}$ couples the FM moment to net canting at the AFM boundary, whereas the antisymmetric combination $J_\Gamma^{(-)}$ couples it directly to the N\'eel texture. Compensated and uncompensated FM/AFM interfaces therefore enter the theory differently already at the level of the variational coupling.

The interface conditions are now coupled. On the FM side one finds
\begin{equation}
\bm J_{\rm FM}
+
\Big(\frac{\delta w_\Gamma}{\delta \bm m}\Big)_{\!\perp}
=0,
\label{eq:FMAFM_m_bc}
\end{equation}
where
\begin{equation}
\bm J_{\rm FM}
=
2A_{\rm FM}\,\partial_n\bm m
+
D_{\rm FM}(\hat{\bm z}\times\hat{\bm n})\times\bm m.
\label{eq:J_FM_mixed}
\end{equation}
On the AFM side, one obtains one boundary condition for each sublattice,
\begin{equation}
\bm J_{\rm AFM}
-
\Big(\frac{\delta w_\Gamma}{\delta \bm m_a}\Big)_{\!\perp}
=0,
\qquad a=1,2,
\label{eq:FMAFM_ma_bc}
\end{equation}
with $\bm J_{\rm AFM}$ given by Equation~\eqref{eq:AFM_flux}. The AFM--FM case is obtained by exchanging left and right. Formally this amounts only to relabelling the two sides, but physically it is a distinct transmission problem because the source and receiving media are interchanged across a boundary at which the underlying order parameter changes.

\subsection{Bulk scales and reduced control parameters}

The interface conditions described in the above subsections specify how the fields must match at the boundary, but successful transmission also requires the receiving medium to support a compact skyrmion texture as it crosses the interface. Relevant bulk scales can be introduced without reference to a particular magnetic order. For a chiral magnetic medium the exchange--anisotropy wall width is
\begin{equation}
\Delta
=
\sqrt{\frac{A}{K_{\mathrm{eff}}}},
\label{eq:Delta}
\end{equation}
and the corresponding N\'eel-wall line energy is
\begin{equation}
\sigma_{\rm DW}
\approx
4\sqrt{A K_{\mathrm{eff}}}-\pi D .
\label{eq:sigma_DW}
\end{equation}
Here $\Delta$ sets the length over which the skyrmion perimeter can bend or compress, while $\sigma_{\rm DW}$ gives the leading energetic cost of that perimeter~\cite{Rohart2013}. The DMI term lowers the wall energy by selecting a preferred chirality, so increasing $D$ softens the skyrmion boundary and eventually favors extended stripe-like textures.

It is therefore useful to introduce the critical DMI scale
\begin{equation}
D_c
=
\frac{4}{\pi}\sqrt{A K_{\mathrm{eff}}},
\label{eq:Dc}
\end{equation}
defined as the DMI value at which the one-dimensional N\'eel-wall energy vanishes in the continuum estimate. We then define the dimensionless DMI fraction
\begin{equation}
\kappa
=
\frac{D}{D_c}
=
\frac{\pi D}{4\sqrt{A K_{\mathrm{eff}}}} .
\label{eq:kappa}
\end{equation}
The quantity $\kappa$ measures the proximity of the medium to the wall-softening threshold. Small $\kappa$ corresponds to a stiff boundary with weak chiral stabilization, whereas $\kappa\sim1$ indicates a soft wall susceptible to stripe formation. These quantities provide the reduced coordinates used below to interpret whether an interface-induced deformation remains localized as a skyrmion or relaxes into collapse or an extended domain texture~\cite{Bogdanov2001,Rohart2013,Muratov2017}.

For a FM region, Eqs.~\eqref{eq:Delta}--\eqref{eq:kappa} are evaluated directly using the FM material parameters $A$, $K_{\mathrm{eff}}$, and $D$. For an AFM region, the same quantities apply to the individual magnetic sublattices. In the strong-exchange limit, this is equivalently expressed in terms of the effective stiffness, anisotropy, and DMI of the staggered N\'eel texture. One may therefore write
\begin{align}
\Delta_{\text{DW}, n}
&\sim
\sqrt{\frac{A_n}{K_n}}, \nonumber
\\[7pt]
D_{c, n}
&\sim
\frac{4}{\pi}\sqrt{A_n K_n},
\label{eq:AFM_scales}
\\[7pt]
\kappa_n
&=
\frac{D_n}{D_c}, \nonumber
\end{align}
where $A_n$, $K_n$, and $D_n$ denote the corresponding N\'eel-field material coefficients. The exact numerical values are model-dependent, because they depend on how the sublattice exchange, anisotropy, and DMI are coarse-grained into the staggered-order description. The interpretation, however, is unchanged: $\Delta$ measures the local stiffness of the boundary, $\sigma_{\rm DW}$ measures the leading line-energy cost of the chiral wall, and $\kappa$ measures its DMI-driven softness.

For mixed-order interfaces, no single reduced variable is sufficient, because transmission depends simultaneously on the source texture, the stability range of the receiving medium, and the interfacial coupling strengths $J_{\Gamma 1}$ and $J_{\Gamma 2}$. Even so, the receiving-side quantities $\Delta_{\mathrm{DW}, R}$, $\sigma_{{\rm DW},R}$, $D_{c,R}$, and $\kappa_R$ remain useful organizing variables. They are obtained from the general definitions in Eqs.~\eqref{eq:Delta}--\eqref{eq:kappa}, evaluated with the receiving-side FM parameters for a FM receiver or with the corresponding sublattice/N\'eel-field parameters for an AFM receiver as summarized in Equation~\eqref{eq:AFM_scales}. Same-order interfaces are governed primarily by bulk mismatch and chiral torque balance, whereas mixed-order interfaces require those ingredients and an additional interfacial conversion condition encoded in $w_\Gamma$.

\section{Simulation results and transmission maps}

To connect the interfacial model to observable transmission outcomes, we carried out micromagnetic simulations of a magnetic thin film using mumax3~\cite{Vansteenkiste2014} and mumax$^+$~\cite{Moreels2026}. The thin film was divided into two regions by a planar interface $\Gamma$, normal to the $x$-axis, as depicted in Figure~\ref{fig:interface_schematic}. In every simulation, the material parameters of the left-hand region were held fixed at $A=1.5\times10^{-12}{\rm Jm^{-1}}$, $D=3.0\times10^{-3}{\rm Jm^{-2}}$, and $K_u=8.0\times10^{5}{\rm Jm^{-3}}$. A skyrmion was initialized in this fixed region and driven across the interface into the receiving region, where its resulting morphology was classified after interaction with the boundary. Across all four interface configurations, the receiving-side material parameters were varied over the same ranges, with the exchange stiffness spanning $A=3.0\times10^{-12}$ to $3.0\times10^{-11}{\rm Jm^{-1}}$, the DMI strength spanning $D=0.75\times10^{-3}$ to $3.00\times10^{-3}{\rm Jm^{-2}}$, and the uniaxial anisotropy in range $K_u=2.0\times10^{5}$ to $1.0\times10^{6}{\rm Jm^{-3}}$. A driving current was constantly applied along the $x$-direction, with its magnitude varied from $|j|=1.0\times10^{12}$ to $9.0\times10^{12}{\rm Am^{-2}}$.

Using the same morphological classification for FM--FM, AFM--AFM, FM--AFM, and AFM--FM interfaces allows the four geometries to be compared on a common footing. The parameter ranges considered here are chosen to be representative of ultrathin chiral multilayers known to host room-temperature skyrmions. FM skyrmions have been observed in heavy-metal/ferromagnet stacks such as Pt/Co/Ta, Pt/Co/W, Ir/Fe/Co/Pt and related Co-based multilayers, where interfacial spin--orbit coupling provides perpendicular magnetic anisotropy and DMI values commonly of order \(D\sim 0.3\)--\(2\,{\rm mJ\,m^{-2}}\), with effective anisotropies in the \(10^{5}\)--\(10^{6}\,{\rm J\,m^{-3}}\) range and exchange stiffnesses typically of order \(A\sim 5\)--\(15\,{\rm pJ\,m^{-1}}\)~\cite{Woo2016,Soumyanarayanan2017,Lin2018,Alshammari2021}. Other skyrmion-hosting platforms are also relevant, although they map less directly onto this interfacial-DMI thin-film parameterization. These include bulk chiral magnets such as MnSi~\cite{Muhlbauer2009}, FeGe~\cite{Yu2011}, and Cu\(_2\)OSeO\(_3\)~\cite{Seki2012}, where skyrmions arise from bulk DMI. 

AFM skyrmions are less commonly realized as single-phase natural AFMs, but have been stabilized experimentally in synthetic AFMs formed from antiferromagnetically coupled FM layers, including Co-based multilayers coupled through Ru spacers; in such systems the perpendicular anisotropy, DMI and interlayer exchange can be tuned together to stabilize compensated skyrmionic textures at room temperature~\cite{Legrand2020,Juge2022}. Ferrimagnetic or antiferromagnetically coupled systems such as GdFeCo further illustrate how compensation can reduce the skyrmion Hall response while retaining chiral-texture stability~\cite{Woo2018b}. These material examples motivate treating the \(A\), \(D\), and \(K_u\) sweeps in the simulations not as arbitrary parameters, but as idealized cuts through experimentally accessible skyrmion-hosting regimes.

\begin{figure}[t]
\centering
\includegraphics[width=\columnwidth]{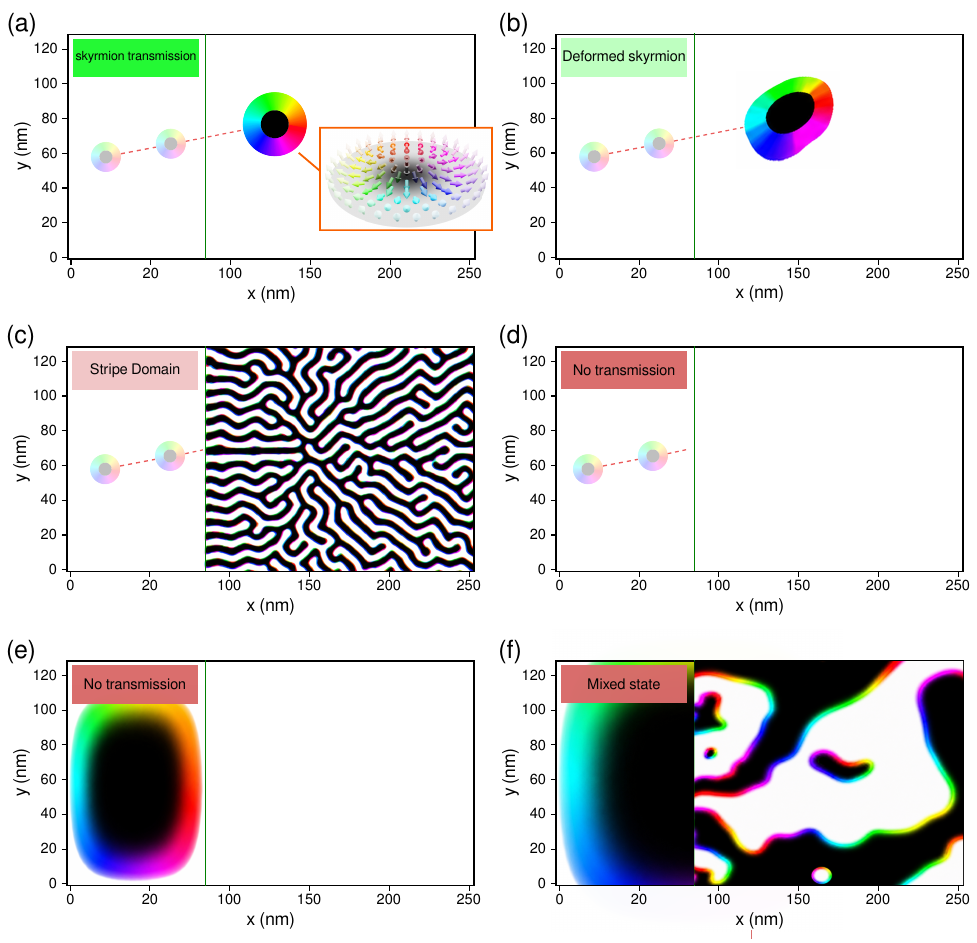}
\caption{\label{fig:morphology_examples}
Representative transmission outcomes, shown using an FM--FM interface as an illustrative example, for the morphological classes used in the phase diagrams. Transparent skyrmions indicate earlier positions along the trajectory indicated by the faded red dashed line, while solid textures show the current or final outcome after interaction with the interface.
(a) Compact skyrmion transmission, in which the incident texture crosses the interface and reforms as a localized skyrmion in the receiving region. The inset in panel (a) shows the spin texture of an isolated skyrmion: the colored arrows represent the local orientation of the magnetic moments, with white and black denoting positive and negative out-of-plane orientations, respectively, and the remaining colors indicating different in-plane orientations.
(b) Deformed skyrmion transmission, where the transmitted object remains skyrmion-like but its perimeter is distorted by the interfacial crossing.
(c) Stripe-domain formation, where the skyrmion boundary opens into an extended domain-wall texture rather than recovering a compact closed shape.
(d) Collapse into the background, where the transmitted texture is lost and the receiving region relaxes to the uniform magnetic state.
(e) Source-side inflation, where the skyrmion remains pinned before the interface and expands into a large bubble-like texture without forming a compact transmitted object.
(f) Transmitted-domain instability, where an inflated source-side texture is partially pushed through the interface and reorganizes in the receiving region into a labyrinthine or stripe-domain network, sometimes containing transient skyrmion-like pockets that may subsequently annihilate.
}
\end{figure}

Before presenting the phase diagrams, it is useful to illustrate the principal outcomes that enter this classification. Figure~\ref{fig:morphology_examples} shows representative examples of the limiting behaviors observed in the simulations, including compact transmission, deformed transmission, stripe-domain formation, and collapse into the background. These examples provide the visual basis for the labels used below, where green regions denote skyrmion-like transmission and regions colored in shades or red denote either stripe-domain formation or loss of the transmitted texture.

\begin{figure}[t]
\centering
\includegraphics[width=\columnwidth]{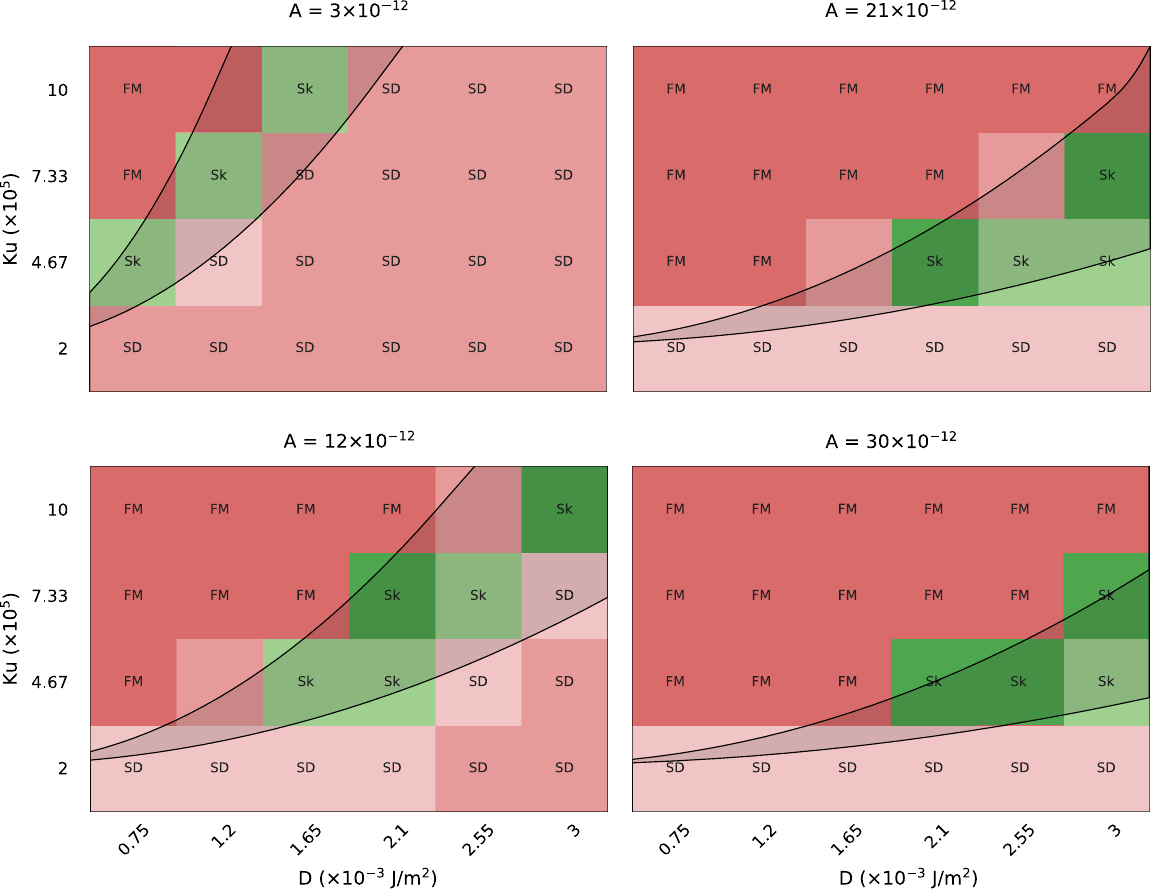}
\caption{\label{fig:fmfm_phase} FM--FM interface phase diagrams showing the structures formed in the variable right-hand region after a skyrmion is driven into it from the fixed left-hand region. Each panel shows the simulated right-region outcome as a function of the right-region DMI strength $D$ and uniaxial anisotropy $K_u$ at a fixed right-region exchange stiffness $A$. All panels correspond to a driving-current magnitude of $|j|=9\times10^{12}{\rm Am^{-2}}$. The left-hand material parameters are held at $A=1.5\times10^{-12}{\rm Jm^{-1}}$, $D=3.0\times10^{-3}{\rm Jm^{-2}}$, and $K_u=8.0\times10^{5}{\rm Jm^{-3}}$. The categorical color scale groups the outcomes into background, stripe-domain, and skyrmion-like morphologies. Points labelled ``FM'' denote relaxation to the FM background. Points labelled ``SD'' denote stripe-domain textures with darker tones corresponding to more extended stripe networks, while lighter tones indicate more compact configurations with some semblance of an amorphous skyrmion-like morphology. Green regions mark skyrmion transmission, with lighter green for slightly deformed skyrmions and darker green for compact transmission. The gray band bounded by black curves gives the analytic \(\kappa\)-corridor as described in Eqs.~\eqref{eq:Dc}~and~\eqref{eq:kappa}, extracted from high-quality transmission points.
}
\end{figure}

The representative outcomes in Figure~\ref{fig:morphology_examples} provide the morphological reference for interpreting the phase maps. The FM--FM phase diagram, Figure~\ref{fig:fmfm_phase}, illustrates the broadest and most continuous set of transmitted morphologies. This behavior is expected for a same-order FM boundary. The skyrmion remains a texture of a single magnetization field on both sides, so the interface primarily tests whether the skyrmion perimeter can adapt to a new exchange--anisotropy--DMI balance. Green regions identify skyrmion transmission, ranging from slightly deformed to compact transmitted skyrmions. Red regions collect outcomes in which compact skyrmion transmission is not obtained. The most chromatically saturated red corresponds to loss of the transmitted texture into the FM background, while the remaining red classes denote stripe-domain morphologies. These stripe domains arise when the core of the incident skyrmion, whose magnetization points opposite to the surrounding uniform background, crosses the interface but loses the radial confinement needed to remain a closed skyrmion. The core then elongates into an extended oppositely magnetized domain, producing a stripe-like morphology. The resulting state may therefore be viewed as skyrmion-seeded magnetization reversal in the receiving ferromagnet. Darker red regions correspond to longer and more connected stripe networks, whereas lighter red regions are more compact and retain some semblance of an amorphous skyrmion-like morphology. 

Further failure modes are shown in Figs.~\ref{fig:morphology_examples}~(e) and (f). In Figure~\ref{fig:morphology_examples} (e), the incident skyrmion remains on the source side of the interface but expands into a large bubble-like texture that occupies much of the left region without crossing the interface. We classify this as failed transmission because no compact skyrmion is established in the receiving medium, even though the source-side texture does not immediately annihilate. A likely origin is interface pinning combined with source-side wall softening. The boundary presents an energetic or dynamical barrier to crossing, while the applied drive continues to press the skyrmion against the interface. If the domain-wall energy on the incident side is sufficiently low, the skyrmion perimeter can expand laterally instead of passing through, producing an inflated, interface-pinned skyrmion-like bubble. Figure~\ref{fig:morphology_examples}~(f) illustrates a more severe transmitted-domain instability. In this case, the skyrmion inflates on the source side as it is driven against the interface, but part of the expanded texture is pushed through into the receiving medium and reorganizes there into an irregular domain-like network. Thus the failure is not simply that the skyrmion remains trapped on the left, nor that it collapses immediately at the boundary. Instead, the interface admits a distorted, over-expanded portion of the texture, which then relaxes on the receiving side into connected reversed domains and internal pockets. Some of these pockets can momentarily resemble isolated skyrmions, but they are not stably transmitted objects and may subsequently annihilate. This behavior is consistent with a receiving medium that can support chiral domain walls but does not provide a restoring energetic balance for a closed skyrmion perimeter. The applied current drive continues to push the magnetic texture across the interface into the receiving region, while the softened domain wall and interfacial mismatch promote lateral expansion, producing a labyrinthine or stripe-domain morphology rather than clean transmission.

The progression from background relaxation to stripe-domain formation and finally to compact transmission reflects the balance between interfacial deformation and the wall energetics of the receiving medium~\cite{Bogdanov2001,Thiaville2012,Rohart2013}. As the skyrmion crosses into a region with different \(A\), \(D\), and \(K_u\), its perimeter must adjust to a new wall width, line tension, and preferred chiral twist. If the receiving-side DMI is too weak relative to exchange and anisotropy, the boundary is too stiff to accommodate the deformation and the texture relaxes toward the uniform state. If the DMI softens the wall too strongly, the perimeter can open into extended stripe-like structures~\cite{Bogdanov2001,Rohart2013,Muratov2017}. Compact transmission therefore occurs in an intermediate regime in which the receiving medium admits the skyrmion without allowing its boundary to expand into a stripe network.

The analytic envelopes overlaid on Figure~\ref{fig:fmfm_phase} are obtained from the reduced DMI fraction of Equation~\eqref{eq:kappa}, as detailed in Appendix~\ref{app:analytic_band}. Once a transmission corridor \(\kappa\in[\kappa_{\min},\kappa_{\max}]\) is extracted from the high-quality transmitted points, the corresponding boundaries in the \((D,K_u)\) plane follow as
\begin{equation}
D_{\mathrm{low/high}}(K_u)
=
\kappa_{\min/\max}\frac{4}{\pi}
\sqrt{A K_{\mathrm{eff}}(K_u)} .
\label{eq:DKu_results_band}
\end{equation}
Only the numerical values of \(\kappa_{\min}\) and \(\kappa_{\max}\) are inferred from the discrete simulations. The envelope curves themselves are analytic consequences of the micromagnetic stability scale \(D_c(A,K_{\mathrm{eff}})\). In the FM--FM case, the envelope captures the central organization of the phase diagram. Compact transmission is concentrated within an intermediate-\(\kappa\) corridor, whereas the low-\(\kappa\) and high-\(\kappa\) sides favor collapse-like and stripe-like relaxation, respectively.

\begin{figure}[t]
\centering
\includegraphics[width=\columnwidth]{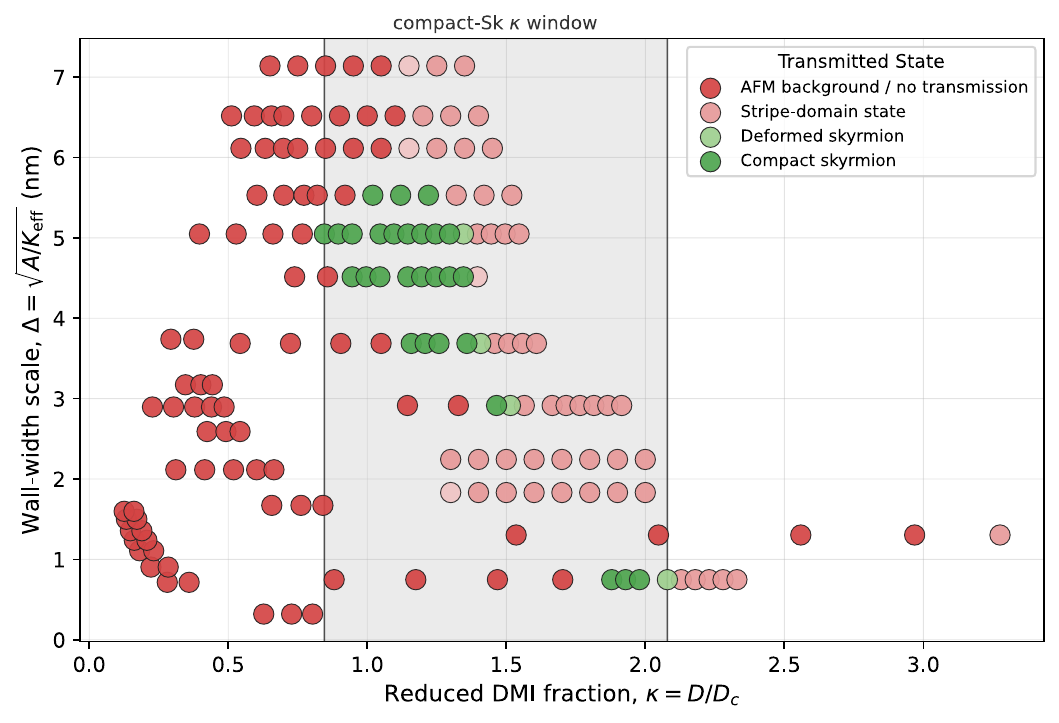}
\caption{\label{fig:afmafm_reduced}
Reduced-coordinate summary of the AFM--AFM interface simulations. Each point corresponds to one receiving-side material parameter set after the outcomes obtained over the sampled drives have been combined, and is plotted using the wall-width scale \(\Delta\) defined in Equation~\eqref{eq:Delta} and the reduced DMI fraction \(\kappa\) defined in Equation~\eqref{eq:kappa}. Color distinguishes relaxation into the AFM background or failure to transmit, stripe-domain formation, deformed-skyrmion transmission, and compact-skyrmion transmission. The shaded vertical band spans the range of \(\kappa\) over which deformed or compact skyrmion reception is observed.
}
\end{figure}

To compare the remaining interface classes without reproducing the full set of phase-diagram grids, we recast the simulation outcomes in the reduced coordinates introduced in Eqs.~\eqref{eq:Delta}--\eqref{eq:kappa}. This representation collapses the separate \(A\), \(D\), and \(K_u\) sweeps onto the wall-width scale \(\Delta\) of Equation~\eqref{eq:Delta} and the reduced DMI parameter \(\kappa\) of Equation~\eqref{eq:kappa}. It therefore organizes the outcomes according to the principal receiving-side energetic scales governing the size, deformability, and stability of the transmitted texture. Figure~\ref{fig:afmafm_reduced} shows this construction for the AFM--AFM interface, where the magnetic order remains AFM across the boundary and transmission does not require conversion between distinct order-parameter manifolds.

The bounded distribution of transmitted morphologies follows from the competition encoded by the domain-wall energy in Equation~\eqref{eq:sigma_DW}. At small \(\kappa\), the DMI-induced reduction of the wall energy is insufficient to support the finite wall length of a transmitted skyrmion. The texture therefore remains subject to a strong contraction tendency and may collapse, become trapped at the interface, or fail to enter the receiving region. Increasing \(\kappa\) lowers the energetic cost of the chiral wall and allows a finite-radius texture to remain metastable after crossing. At still larger \(\kappa\), however, the cost of creating additional wall length becomes sufficiently small that a closed skyrmion wall is no longer favored over an extended configuration. Radial confinement is then weakened and the transmitted texture evolves into a deformed-skyrmion or stripe-domain morphology. Compact transmission is consequently restricted by both a lower bound associated with collapse or interfacial trapping and an upper bound associated with the loss of radial confinement and stripe proliferation.

Because both bounds depend on the wall-width scale, the transmission condition is more appropriately expressed as
\begin{equation}
\kappa_{\min}(\Delta)
<
\kappa
<
\kappa_{\max}(\Delta),
\label{eq:dynamic_transmission_window}
\end{equation}
rather than as a universal interval of \(\kappa\). The broadest compact-transmission region occurs at intermediate \(\Delta\), where the texture is sufficiently compliant to adjust to the material discontinuity while retaining enough radial confinement to remain closed. At larger \(\Delta\), the wall is broader and more deformable, so increasing the DMI more readily promotes expansion into stripe-like configurations. At smaller \(\Delta\), the stronger confinement favors contraction and increases the DMI required to prevent collapse. The compact-transmission pocket below \(\Delta=1~\mathrm{nm}\) is consistent with this latter regime. Strong anisotropy suppresses broad domain expansion, while sufficiently large DMI can preserve a narrow chiral core. Since this wall-width scale is comparable to the spatial discretization, however, the precise extent of this regime is expected to be more mesh-sensitive than the intermediate-\(\Delta\) transmission region.

Although FM--FM and AFM--AFM are both same-order geometries, preservation of the magnetic-order type does not guarantee identical transmission behavior. At an FM--FM interface, material mismatch can be accommodated through continuous deformation of a single magnetization field. In the AFM--AFM case, the corresponding deformation must preserve the antiparallel relation between the two sublattices and maintain coherence of the N\'eel order. This additional constraint reduces the range of permissible reconstructed textures and produces a comparatively sharp separation among relaxation into the background, stripe-domain formation, and compact-skyrmion transmission. The distribution in Figure~\ref{fig:afmafm_reduced} should therefore be interpreted as a dynamical transmission window rather than an equilibrium phase boundary, because the observed state depends not only on the wall-width scale of Equation~\eqref{eq:Delta}, the domain-wall energy of Equation~\eqref{eq:sigma_DW}, and the reduced DMI parameter of Equation~\eqref{eq:kappa}, but also on the interfacial barrier, the applied drive, and the finite geometry.

The distinction is also dynamical. In FM media, a driven skyrmion generally acquires a transverse velocity through its gyrotropic, or Magnus-like, response. This skyrmion Hall effect deflects the trajectory away from the direction of the applied drive. In an interface geometry, such transverse motion changes the position and angle at which the skyrmion encounters the material boundary and can enhance edge interactions, local pinning, and asymmetric deformation. In an ideal AFM, by contrast, the opposite gyrotropic responses of the two sublattices largely compensate, suppressing the conventional skyrmion Hall deflection. The absence of a strong transverse drift helps organize the AFM--AFM outcomes more sharply in the reduced coordinates, although transmission remains constrained by the requirement that the two sublattice textures cross the interface coherently.

\begin{figure}[t]
\centering
\includegraphics[width=\columnwidth]{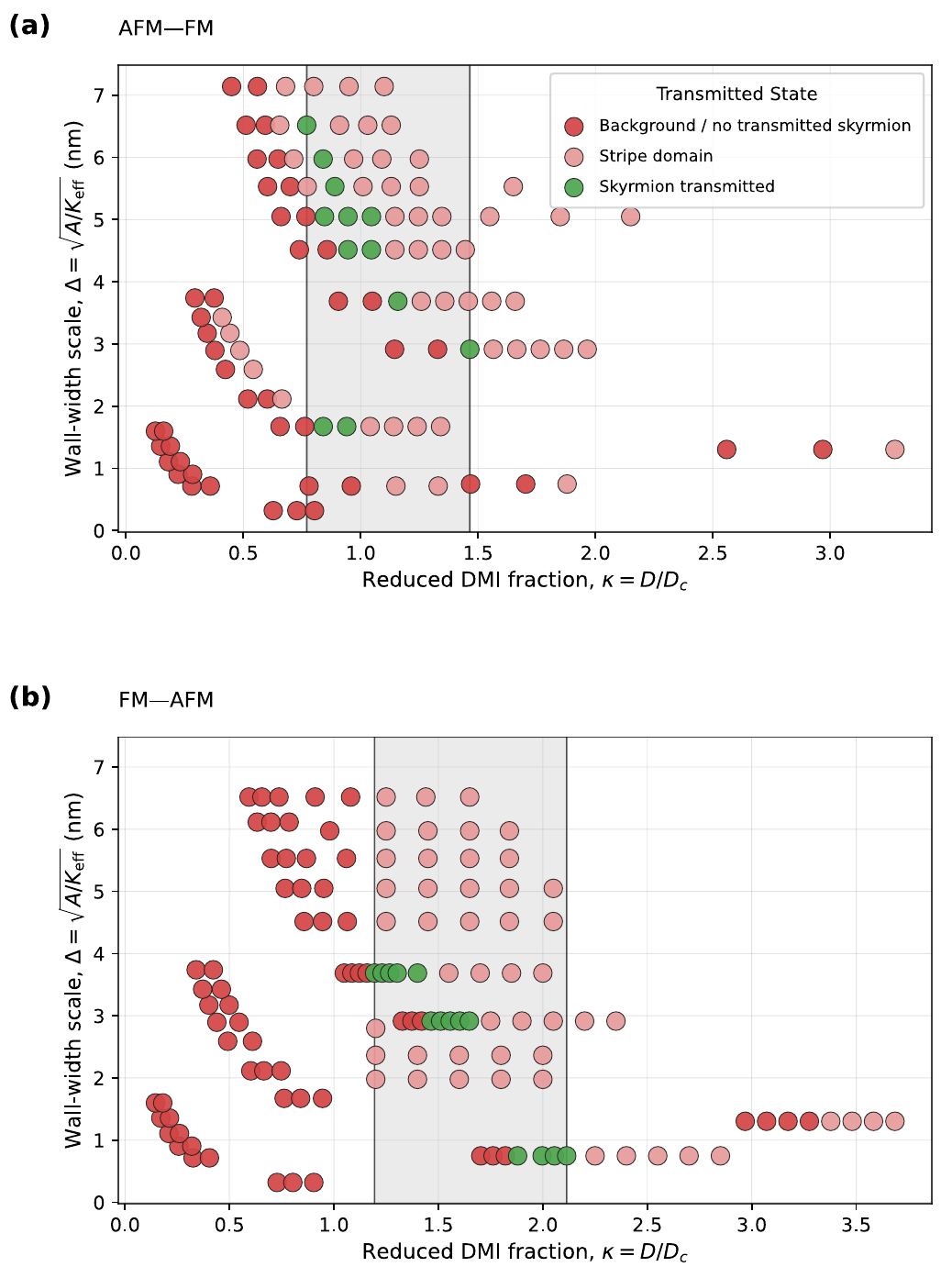}
\caption{\label{fig:mixed_order_reduced}
Reduced-coordinate comparison of mixed-order skyrmion transmission for (a) AFM--FM and (b) FM--AFM interfaces. Each point represents one receiving-side material parameter set after combining the outcomes obtained over the sampled drive values, and is plotted using the wall-width scale \(\Delta\) defined in Equation~\eqref{eq:Delta} and the reduced DMI fraction \(\kappa\) defined in Equation~\eqref{eq:kappa}. Color distinguishes relaxation into the background or failure to transmit, stripe-domain formation, and skyrmion transmission. The shaded bands span the values of \(\kappa\) associated with skyrmion transmission in each geometry. The different distributions in panels (a) and (b) demonstrate that AFM--FM and FM--AFM transmission are directionally distinct order-conversion processes rather than equivalent reversals of the same interface.
}
\end{figure}

In AFM--FM and FM--AFM geometries, the boundary is not merely a material discontinuity within a common order-parameter manifold; it must mediate conversion between an FM magnetization texture and an AFM staggered texture. The wall-energy competition established above continues to delimit collapse-dominated behavior at insufficient DMI and stripe-domain formation once radial confinement is lost, but conversion of the magnetic order imposes an additional constraint on the interval over which a compact transmitted texture can survive. Figure~\ref{fig:mixed_order_reduced} shows that this interval depends strongly on the propagation direction. In the AFM--FM case, transmission extends over a comparatively broad range of wall-width scales and occurs predominantly at moderate reduced DMI, whereas the FM--AFM transmission region is concentrated into fewer branches and is generally displaced toward larger \(\kappa\). The receiving-side wall scales therefore remain necessary descriptors of the transmitted-state stability, but they do not determine the outcome independently of the conversion pathway and the accompanying reconstruction of the magnetic order.

In the AFM--FM geometry shown in Figure~\ref{fig:mixed_order_reduced}(a), the compensated two-sublattice texture must evolve into an effectively single-sublattice FM texture. Skyrmion transmission is observed over wall-width scales ranging from approximately \(\Delta\approx1.7~\mathrm{nm}\) to above \(6~\mathrm{nm}\), with most transmitted states occurring near \(\kappa\approx0.8\)--\(1.2\). The required reduced DMI shifts toward larger values for some narrower-wall branches, indicating that stronger chiral stabilization is needed as the receiving texture becomes more tightly confined. The comparatively broad distribution in \(\Delta\) suggests that the FM receiving medium can accommodate the loss of one sublattice component through several reconstructed configurations. Nevertheless, much of the parameter space surrounding the compact-transmission region evolves into extended or stripe-like morphologies, showing that successful conversion does not necessarily preserve radial compactness. Once conversion has occurred, the transmitted FM texture also acquires a gyrotropic response, so transverse motion can provide an additional channel for asymmetric deformation and displacement within the receiving region.

The reverse FM--AFM process shown in Figure~\ref{fig:mixed_order_reduced}(b) is more spatially restricted in the reduced-coordinate plane. The principal compact-transmission branches occur near \(\Delta\approx2.9\)--\(3.7~\mathrm{nm}\) and are shifted toward approximately \(\kappa\approx1.2\)--\(1.7\). A separate narrow-wall branch appears below \(\Delta=1~\mathrm{nm}\), where transmission requires still larger reduced DMI, around \(\kappa\approx1.8\)--\(2.1\). By contrast, no compact transmitted states are observed over the larger wall widths sampled in this geometry, where the response is dominated by non-transmission or stripe-domain formation. This restricted distribution is consistent with the reconstruction required on entering the AFM. The incident FM skyrmion carries a single magnetization texture, while the receiving medium must generate a second sublattice with the appropriate antiparallel alignment and spatial registration to form a coherent compensated pair. The resulting sublattice-registration cost couples translational passage through the interface to the internal reconstruction of the texture, so a metastable skyrmion supported by the receiving AFM is not necessarily reachable dynamically from the incident FM configuration.

The contrast between Figs.~\ref{fig:mixed_order_reduced}(a) and \ref{fig:mixed_order_reduced}(b) therefore reflects more than a reversal of the propagation direction. AFM--FM transmission removes one component of the staggered order and produces an FM texture over a relatively broad range of wall widths, whereas FM--AFM transmission must construct a compensated two-sublattice texture and is consequently confined to narrower intervals of \(\Delta\) and generally larger values of \(\kappa\). The observed distributions are thus governed jointly by the wall-width scale in Equation~\eqref{eq:Delta}, the domain-wall energy in Equation~\eqref{eq:sigma_DW}, the reduced DMI parameter in Equation~\eqref{eq:kappa}, the translational interface barrier, and the direction-dependent reconstruction of the magnetic order.

Taken together, the phase diagrams and reduced summaries show that planar interfaces act as active elements of skyrmion transport rather than passive material discontinuities. FM--FM transmission is governed primarily by perimeter deformation, receiving-side wall energetics, and the transverse dynamics characteristic of FM skyrmions. It is therefore well organized by the receiving-side \(\kappa\)-corridor when the Hall-induced trajectory remains compatible with boundary crossing. AFM--AFM transmission preserves the order type but imposes a stricter coherence condition on the staggered texture, while also suppressing the conventional skyrmion Hall deflection through sublattice compensation. Mixed-order interfaces add an order-parameter conversion requirement, making AFM--FM and FM--AFM directionally distinct rather than interchangeable reversals of the same boundary. The combined analysis therefore identifies the relevant design variables for heterogeneous skyrmion tracks. Bulk skyrmion viability, interfacial deformation, order-parameter conversion, and Hall-induced dynamical asymmetry.

Our work is has important implications because it converts lateral interface transmission from a case-by-case numerical observation into a materials-design problem. Rather than asking only whether a skyrmion is stable in each homogeneous region, we show that one can ask whether a chosen boundary places the receiving medium in the appropriate reduced stability window and whether the interface geometry preserves, converts, or destabilizes the relevant order parameter. This framework is directly relevant for heterogeneous skyrmion tracks, where nucleation, transport, storage, readout, and deletion may be optimized in different material regions. In this setting, an interface need not be treated only as a source of loss. It can be designed as an active element that filters, selects, converts, or resets chiral magnetic textures, and may provide a route to skyrmion diodes~\cite{Jung2021,Yan2026}, rectifiers~\cite{Lin2013}, and ratchet-like transport under time-dependent drives~\cite{Reichhardt2015,Gobel2021}.

These predictions could be tested experimentally in patterned multilayers where the material parameters or magnetic order are varied laterally across a controlled boundary. One direct route would be to fabricate a track in which a conventional FM skyrmion-hosting multilayer continues into a synthetic AFM, for example by adding or activating a second antiferromagnetically coupled FM layer across part of the device. Such a geometry would realize an effective FM--AFM interface within a continuous transport channel. The source side would support an ordinary FM skyrmion, while the receiving side would favor a compensated skyrmion pair.

A natural extension of this framework is to broader families of magnetic textures~\cite{Gobel2021b} and interface geometries. Beyond skyrmions, one can examine how merons~\cite{Ezawa2011}, antiskyrmions~\cite{Koshibae2016}, bimerons~\cite{Kharkov2017}, and skyrmionium~\cite{Zhang2016b,Gobel2019} textures transmit, deform, or convert at the same class of boundaries. The framework could also be extended to altermagnets, which combine compensated magnetic order with a spin-split electronic structure and therefore provide an interesting case beyond the interfaces considered here. Other natural directions include graded rather than abrupt material changes, disorder, edge roughness, different interfacial exchange conditions, curved or circular interfaces~\cite{Muller2015}, and periodic arrays of one- or two-dimensional interfaces~\cite{Reichhardt2022,Saha2019,Souza2024}. In such settings, steady-state motion across repeated boundaries could lead to periodic breathing, size modulation, or morphology conversion, while skyrmions stabilized directly on an interface could form Janus skyrmions~\cite{Zhang2026}. More generally, combining reduced-coordinate phase maps with interface-specific coupling parameters offers a route toward predictive design rules for modular skyrmion circuitry, where interfaces are deliberately chosen to transmit, reshape, rectify, or eliminate skyrmionic information carriers.

\section{Acknowledgments}
We appreciate the support of the University of California LFRP grant L25CR8980. We also gratefully acknowledge that the work at Los Alamos National Laboratory was carried out under the auspices of the U.S. Department of Energy and National Nuclear Security Administration under Contract No. 89233218CNA000001. 

\appendix

\setcounter{section}{0}
\renewcommand{\thesection}{\Alph{section}}
\refstepcounter{section}
\label{app:interface_variation}

\section*{Appendix \thesection: Variational origin of the interface torque-balance conditions}

Here we present the variational derivation of the interface conditions used in the main text. The derivation is independent of the specific magnetic order on either side of the interface and therefore applies to FM--FM, AFM--AFM, FM--AFM, and AFM--FM geometries, provided the appropriate order parameter is used in each region.

We consider the total energy
\begin{align}
E[\bm q_L,\bm q_R]
=&
\int_{\Omega_L} d^2r\,\mathcal E_L(\bm q_L,\nabla\bm q_L)
+
\int_{\Omega_R} d^2r\,\mathcal E_R(\bm q_R,\nabla\bm q_R)
\nonumber
\\[2mm]
&+
\int_\Gamma ds\,w_\Gamma(\bm q_L,\bm q_R),
\label{eq:app_Etot}
\end{align}
where \(\Omega_L=\{x<0\}\), \(\Omega_R=\{x>0\}\), and \(\Gamma=\{x=0\}\). The unit normal \(\hat{\bm n}\) is chosen to point from \(L\) to \(R\). The outward normal of \(\Omega_L\) at the interface is therefore \(\hat{\bm n}\), whereas the outward normal of \(\Omega_R\) at the same interface is \(-\hat{\bm n}\). The fields \(\bm q_L\) and \(\bm q_R\) denote the appropriate order parameters on the two sides. For a FM \(\bm q=\bm m\), while for an AFM \(\bm q\) may denote the N\'eel vector $\bm n$, the sublattice magnetizations, or another constrained set of variables.

For each bulk region, the first variation of the energy has the form
\begin{align}
\delta E_\alpha
=&
\int_{\Omega_\alpha} d^2r\,
\left[
\frac{\partial \mathcal E_\alpha}{\partial \bm q_\alpha}
-
\partial_i
\left(
\frac{\partial \mathcal E_\alpha}
{\partial(\partial_i\bm q_\alpha)}
\right)
\right]\cdot\delta\bm q_\alpha \nonumber \\[2mm]
&+ \int_{\partial\Omega_\alpha} ds\,
\left[
\frac{\partial \mathcal E_\alpha}
{\partial(\nabla\bm q_\alpha)}
\cdot \hat{\bm n}_\alpha
\right]\cdot\delta\bm q_\alpha ,
\label{eq:app_bulk_variation}
\end{align}
where \(\alpha\in\{L,R\}\), \(\hat{\bm n}_\alpha\) is the outward normal of \(\Omega_\alpha\), and repeated spatial indices are summed. The first term gives the Euler--Lagrange equations in the interior of each region. The second term is localized on the boundary and gives the natural boundary condition. At the internal interface, the two boundary contributions have opposite signs because the outward normals of the two regions are opposite.

It is useful to define the boundary fluxes with respect to the single interface normal \(\hat{\bm n}\):
\begin{equation}
\bm J_\alpha
=
\left[
\frac{\partial \mathcal E_\alpha}
{\partial(\nabla\bm q_\alpha)}
\cdot \hat{\bm n}
\right]_{\Gamma}.
\label{eq:app_boundary_flux}
\end{equation}
With this convention, the interface-localized variation of the two bulk terms is
\begin{equation}
\delta E_{\rm bulk}\big|_\Gamma
=
\int_\Gamma ds\,
\left[
\bm J_L\cdot\delta\bm q_L
-
\bm J_R\cdot\delta\bm q_R
\right].
\label{eq:app_bulk_boundary_terms}
\end{equation}
The minus sign multiplying the right-side contribution follows from the fact that the outward normal of \(\Omega_R\) is \(-\hat{\bm n}\), while \(\bm J_R\) has been defined using \(+\hat{\bm n}\).

The first variation of the explicit interface energy is
\begin{equation}
\delta E_\Gamma
=
\int_\Gamma ds\,
\left[
\frac{\delta w_\Gamma}{\delta\bm q_L}\cdot\delta\bm q_L
+
\frac{\delta w_\Gamma}{\delta\bm q_R}\cdot\delta\bm q_R
\right].
\label{eq:app_interface_variation}
\end{equation}
Combining Eqs.~\eqref{eq:app_bulk_boundary_terms} and \eqref{eq:app_interface_variation}, the total variation localized on \(\Gamma\) is
\begin{align}
\delta E\big|_\Gamma
=&
\int_\Gamma ds\,
\bigg[
\left(
\bm J_L
+
\frac{\delta w_\Gamma}{\delta\bm q_L}
\right)\cdot\delta\bm q_L \nonumber \\[2mm]
&+ \left(
-\bm J_R
+
\frac{\delta w_\Gamma}{\delta\bm q_R}
\right)\cdot\delta\bm q_R
\bigg].
\label{eq:app_total_boundary_variation}
\end{align}
Stationarity requires this expression to vanish for all admissible variations at the interface. If the order parameters were unconstrained, the coefficients of \(\delta\bm q_L\) and \(\delta\bm q_R\) would vanish directly. Magnetic order parameters, however, are usually constrained. For example, a FM magnetization satisfies \(|\bm m|=1\), and an AFM N\'eel vector may similarly satisfy \(|\bm n|=1\). The variations are therefore restricted to the tangent space of the constraint manifold. Only the component of the variational force tangent to that manifold contributes to the stationarity condition.

Denoting projection onto the allowed tangent space by the subscript \(\perp\), the natural interface conditions are
\begin{equation}
\bm J_L
+
\left(
\frac{\delta w_\Gamma}{\delta\bm q_L}
\right)_{\perp}
=0,
\qquad
\bm J_R
-
\left(
\frac{\delta w_\Gamma}{\delta\bm q_R}
\right)_{\perp}
=0.
\label{eq:app_interface_balance}
\end{equation}
These are the interface torque-balance conditions quoted in the main text. They state that the boundary flux generated by the bulk micromagnetic energy on each side must be balanced by the torque arising from the explicit interfacial coupling energy. The opposite signs on the two sides are a consequence of the opposite orientations of the two outward normals at an internal boundary.

For a unit-vector FM order parameter, the tangent projection of any vector \(\bm a\) may be written as
\begin{equation}
\bm a_\perp
=
\bm a-(\bm a\cdot\bm m)\bm m
=
\bm m\times(\bm a\times\bm m).
\label{eq:app_projection}
\end{equation}
Thus the same condition can equivalently be written in torque form by taking the cross product with the local order parameter. For instance, on a FM side,
\begin{equation}
\bm m_\alpha\times
\left[
\bm J_\alpha
\pm
\frac{\delta w_\Gamma}{\delta\bm m_\alpha}
\right]
=0,
\label{eq:app_torque_form}
\end{equation}
with the sign chosen according to Equation~\eqref{eq:app_interface_balance}. This form emphasizes that the interface condition fixes the balance of torques rather than the longitudinal component of the variational derivative, which is absorbed by the constraint enforcing \(|\bm m_\alpha|=1\).

For a plain material step within a single magnetic order, one may set \(w_\Gamma=0\) if no additional exchange or anisotropy is assigned to the boundary itself. The interface condition then reduces to continuity of the appropriate micromagnetic boundary flux across the step, together with any continuity condition imposed on the order parameter. For mixed-order interfaces, such as FM--AFM or AFM--FM boundaries, \(w_\Gamma\) is essential because the two sides are described by distinct order parameters. The interfacial energy then supplies the conversion torque that couples the FM and AFM textures at \(\Gamma\).

\setcounter{section}{1}
\renewcommand{\thesection}{\Alph{section}}

\refstepcounter{section}
\label{app:analytic_band}
\section*{Appendix \thesection: Analytic envelope curves from the DMI fraction}

The main text overlays analytic corridors on the simulation phase diagrams to indicate where the receiving medium lies within the reduced DMI range associated with compact transmission. For each simulated material point we compute
\begin{equation}
\kappa
=
\frac{D}{D_c}
=
\frac{\pi D}{4\sqrt{A K_{\mathrm{eff}}}},
\label{eq:kappa_repeat_general}
\end{equation}
where \(D_c=(4/\pi)\sqrt{A K_{\mathrm{eff}}}\). The corridor endpoints \(\kappa_{\min}\) and \(\kappa_{\max}\) are extracted from the subset of points classified as successful transmission, for example compact or near-compact transmitted skyrmions, by taking the lower and upper \(\kappa\) values within that subset. For each two-dimensional phase diagram, the parameter not plotted on the axes is fixed by the panel choice, and the corridor curves are obtained by solving Equation~\eqref{eq:kappa_repeat_general} for the parameters plotted on the axes.

For a transmission window
\begin{equation}
\kappa\in[\kappa_{\min},\kappa_{\max}],
\end{equation}
and with \(D\) as the dependent variable, the lower and upper envelopes are
\begin{equation}
D_{\mathrm{low/high}}
=
\kappa_{\min/\max}\,\frac{4}{\pi}\sqrt{A K_{\mathrm{eff}}}.
\label{eq:D_general_band_revised}
\end{equation}
At fixed \(A\), this gives the boundaries in the \((D,K_u)\) plane,
\begin{equation}
D_{\mathrm{low/high}}(K_u)
=
\kappa_{\min/\max}\,\frac{4}{\pi}
\sqrt{A\,K_{\mathrm{eff}}(K_u)}.
\label{eq:DKu_band_revised}
\end{equation}
At fixed \(K_u\), the corresponding boundaries in the \((A,D)\) plane are
\begin{equation}
D_{\mathrm{low/high}}(A)
=
\kappa_{\min/\max}\,\frac{4}{\pi}
\sqrt{A K_{\mathrm{eff}}}.
\label{eq:AD_band_revised}
\end{equation}
For phase diagrams in the \((A,K_u)\) plane at fixed \(D\), Equation~\eqref{eq:kappa_repeat_general} is instead inverted to give
\begin{equation}
A_{\mathrm{low/high}}(K_u)
=
\frac{1}{K_{\mathrm{eff}}(K_u)}
\left(
\frac{\pi D}{4\kappa_{\max/\min}}
\right)^2.
\label{eq:AKu_band_revised}
\end{equation}
The reversed ordering in Equation~\eqref{eq:AKu_band_revised} follows from \(A\propto\kappa^{-2}\) at fixed \(D\). Thus, the different apparent envelope shapes in the various projections are not distinct criteria, but rearrangements of the same reduced scale \(D_c(A,K_{\mathrm{eff}})\).

For mixed-order interfaces, the same receiving-side construction remains useful when the interfacial coupling sector is held fixed. In that case one may define a receiving-side corridor in terms of \(\kappa_R\), but the numerical values of \(\kappa_{\min}\) and \(\kappa_{\max}\) become conditional on the conversion physics encoded in \(w_\Gamma\). In practice, the plotted band may be defined either as an \emph{envelope band}, using the full extremal range of \(\kappa\) among transmitted points, or as a \emph{robust band}, using quantiles of the same subset. The analytic forms remain Eqs.~\eqref{eq:DKu_band_revised}--\eqref{eq:AKu_band_revised}, only the selected numerical endpoints of the corridor change. 

\bibliographystyle{apsrev4-2}
\bibliography{references}

\end{document}